\documentclass[twocolumn,showpacs,prb,citeautoscript]{revtex4}
\usepackage{graphicx}
\usepackage{dcolumn}
\usepackage{bm}
\usepackage{amsmath}
\usepackage{amssymb}
\usepackage{epsfig}
\usepackage{array}
\begin{document}
\title{The two colors of MgB$_{2}$}
\author{V.\ Guritanu }
\author{A.B.\ Kuzmenko}
\author{D.\ van der Marel}
\affiliation{DPMC, University of Geneva, 24, Quai E.-Ansermet,
1211 Geneva 4, Switzerland}
\author{S.M.\ Kazakov}
\author{N.D.\ Zhigadlo}
\author{J.\ Karpinski}
\affiliation{Solid State Physics Laboratory, ETH 8093 Zurich,
Switzerland}
\date{\today }

\begin{abstract}
We present the anisotropic optical conductivity of MgB$_{2}$
between 0.1 and 3.7 eV at room temperature obtained on single
crystals of different purity by the spectroscopic ellipsometry and
reflectance measurements. The bare (unscreened) plasma frequency
$\omega_{p}$ is almost isotropic and equal to 6.3 eV, which
contrasts some earlier reports of a very small value of
$\omega_{p}$. The data suggests that the $\sigma$-bands are
characterized by a stronger electron-phonon coupling $\lambda_
{tr}$ but smaller impurity scattering $\gamma_{imp}$, compared to
the $\pi$-bands. The optical response along the boron planes is
marked by an intense interband transition at 2.6 eV, due to which
the reflectivity plasma edges along the a- and c-axes are shifted
with respect to each other. As a result, the sample spectacularly
changes color from a blueish-silver to the yellow as the
polarization is rotated from the in-plane direction towards the
c-axis. The optical spectra are in good agreement with the
published {\it ab initio} calculations. The remaining
discrepancies can be explained by the relative shift of
$\sigma$-bands and $\pi$-bands by about 0.2 eV compared to the
theoretical band structure, in agreement with the de Haas-van
Alphen experiments. The widths of the Drude and the interband
peaks are both very sensitive to the sample purity.
\end{abstract}

\maketitle
\section{Introduction}

The discovery of superconductivity in MgB$_{2}$
\cite{NagamatsuNature01} caused an excitement in the physics
community not only due to an unprecedentedly high $T_{c}$
($\sim$40 K) for a 'conventional' electron-phonon superconductor,
but also because it clearly shows the existence of two distinct
gaps\cite{MazinPC03}; an intriguing phenomenon that, although
addressed theoretically, had been never observed before. Magnesium
diboride consists of graphite-like boron planes intercalated with
Mg atoms. The metallic properties are determined by two distinct
types of electronic bands: the strongly covalent almost 2D
$\sigma$-bands formed by the hybridized $sp_{x}p_{y}$ B orbitals
and 3D $\pi$-bands made of $p_{z}$ orbitals. The holes in the
$\sigma$-bands are strongly coupled to the in-plane
bond-stretching phonon modes, giving rise to a high
electron-phonon coupling constant\cite{KongPRB01}. A remarkable
implication of such a strong conduction band disparity is a
multigap superconductivity: a large gap value on $\sigma$-bands
and a small one on $\pi$-bands. Notably, the community has enjoyed
a rapid advance in understanding the electronic structure and
superconducting scenario of MgB$_{2}$, in contrast to the lengthy
siege of the high-$T_{c}$ problem in the cuprates.

The early theoretical predictions of the electronic structure and
the superconducting properties were soon confirmed by the isotope
effect \cite{BudkoPRL01,HinksNature01}, angle-resolved
photoemission\cite{UchiyamaPRL02}, de Haas-van Alphen (dHvA)
\cite{YellandPRL02}, specific heat
\cite{WangCM02,BouquetCM02,GolubovJPhysCM02}, tunneling
measurements \cite{EskildsenPRL02} and, recently, by the inelastic
X-ray scattering \cite{GalambosiPRB05}. Even though the
far-infrared
experiments\cite{GorshunovEPJB01,KaindlPRL02,PimenovPRB02,JungPRB02,PerucchiPRL02,OrtolaniCM04}
have clearly shown the lowest of the superconducting gaps in
agreement with the theory and other spectroscopic probes, the
optical measurements so far demonstrated a rather poor
reproducibility and equally poor consistency with the theoretical
electronic structure. The most controversial issue is value of the
Drude plasma frequency $\omega_{p}$. While the theory predicts a
high value of $\omega_{p}$ of $\approx$ 7 eV, which corresponds to
about 1 conducting electron per unit cell, a number of groups
\cite{TuPRL01,KaindlPRL02,KuzmenkoSSC02,MazinPRL02,MunJS02,FudamotoPRB03}
reported a much smaller value of about 1.5 - 2.5 eV, corresponding
to 0.15 electrons per cell. In Ref.\onlinecite{KuzmenkoSSC02}, in
addition to the narrow Drude peak, a broad continuum, which could
contain the missing Drude spectral weight, was found below 1 eV.
However, the existence of such a continuum was not reliably
verified by other groups. Another inconsistency is related to the
anisotropy of the plasma frequency. The calculations provide very
close (within 5\%) values of $\omega_{p,a}$ and $\omega_{p,c}$,
which should be regarded, in fact, as a coincidence, because the
plasma frequency of the carriers in the 2D $\sigma$-bands is
strongly anisotropic. A study of the optical anisotropy of
magnesium diboride was undertaken by Fudamoto and
Lee\cite{FudamotoPRB03} by the comparison of the reflectivity
spectra measured on a mosaic of ab-oriented crystals and on a
polycrystalline sample. They observed the in-plane reflectivity
plasma edge at about 16000 cm$^{-1}$. The authors also suggested
that an additional structure in the reflection of a polycrystal is
due to the c-axis plasma edge at 22000 cm$^{-1}$. Thus the
anisotropy ratio $\omega_{p,a}/\omega_{p,c}\approx 0.73$ was
deduced, which is in contrast with the theoretical prediction.

The optically derived electron-phonon coupling constant
$\lambda_{tr}$ was also a subject of
debate\cite{TuPRL01,MunJS02,MarsiglioPRL01,MaksimovPRL02}. Values
of $\lambda_{tr}$ ranging from 0.13\cite{TuPRL01} to 1.5
\cite{MunJS02} were reported. It was pointed out in
Ref.\onlinecite{MaksimovPRL02} that the determination of
$\lambda_{tr}$ relies strongly on the plasma frequency, which is
not yet well established. One should keep in mind that different
values of the coupling constant for the $\sigma$- and $\pi$-bands
are expected from the calculations\cite{KongPRB01} and observed in
the dHvA experiment\cite{YellandPRL02}.

With a lack of large single crystals of MgB$_{2}$, especially
along the c-axis, the optical measurements were done on
polycrystalline
samples\cite{KuzmenkoSSC02,OrtolaniCM04,FudamotoPRB03},
disoriented\cite{MunJS02,ChvostovaTSF04} or
ab-oriented\cite{TuPRL01,JungPRB02,PimenovPRB02} films as well as
ab-oriented crystal mosaics\cite{PerucchiPRL02,FudamotoPRB03}. We
believe that a large spread of the published optical results can
be explained by (i) different purity levels of the samples used,
(ii) experimental difficulties to extract the anisotropic complex
conductivity from the standard measurements on such objects and
(iii) a fast rate of the surface degradation in
air\cite{KuzmenkoSSC02,FudamotoPRB03}.

In this work, we do optical experiments on individual freshly
polished single crystals of different purity having no contact
with the air. The real and imaginary parts of the in-plane and
out-of-plane optical conductivities are derived from the
reflectivity and ellipsometry measurements on the ab- and, for the
first time, on the ac-oriented crystals. The measurements show a
large ($\approx$ 6.3 eV) and almost isotropic plasma frequency. We
find different positions of the reflectivity plasma edges for the
two light polarizations. However, such anisotropy is due to extra
screening of charge carriers by an interband transition at 2.6 eV,
optically active along the ab-plane, rather than the anisotropy of
the bare plasma frequency itself. The anisotropic spectra are in
satisfactory agreement with the first-principles calculations of
the band structure and the electron-phonon interaction. We discuss
the possible corrections to the electronic structure which follow
from the optical results.

Finally, the presented data tell us what is the 'true' color (or
rather, colors) of MgB$_{2}$.

\section{Samples and preparation}

\begin{figure}[bht]
\centerline{\includegraphics[width=7cm,clip=true]{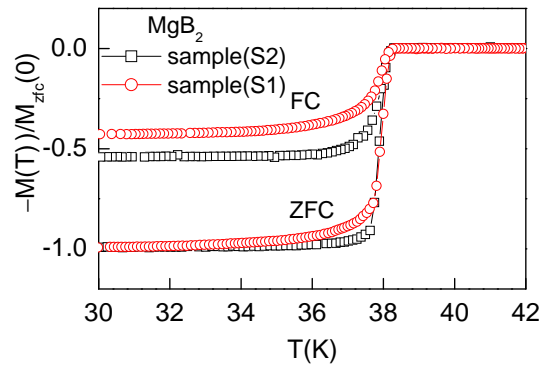}}
\caption{Magnetization measurements of the superconducting
transition of the two single crystals of MgB$_2$ used in this
paper.} \label{FigTrans}
\end{figure}

High-quality single crystals of MgB$_2$ have been grown using a
cubic anvil technique via the peritectic decomposition reaction of
the MgNB$_{9}$ and Mg at temperature up to 2000$^{\circ}$C under a
pressure of 30 - 35 kbar. The details of the crystal growth and
extensive characterization are given elsewhere
\cite{KarpinskiSST02}. The measurements were done on two samples,
referenced below as S1 and S2. The samples were grown under
slightly different conditions. For the sample S1 the purity of Mg
was 99.8\%, while a 4N-pure (99.98\%) magnesium was used in the
second case. The maximum temperature of crystal growth was higher
for the sample S2 by about $80^{\circ}$C. The time at highest
temperature was 10 minutes for the first sample and 30 minutes for
the second one. The $T_{c}$ of both samples is close to 38 K (see
Fig.\ref{FigTrans}), although sample S2 shows somewhat narrower
transition and a larger field-cooled diamagnetic signal compared
to the one of S1.

The dimensions ($a\times b\times c$) of the as-grown crystals were
0.7 $\times$ 0.5 $\times$ 0.27 mm$^{3}$ for S1 and 0.6 $\times$
0.5 $\times$ 0.18 mm$^{3}$ for S2. We selected a thicker sample S1
to prepare the (ac) optical face, while sample S2 was used to
measure on the (ab) face. The faces were dry-polished using a 0.1
$\mu$m diamond abrasive, since the as-grown surfaces were not
suitable for quantitative optical examination.

As it was noticed before\cite{KuzmenkoSSC02,FudamotoPRB03}, the
exposed surface of MgB$_2$ deteriorates quickly, largely due to
the air moisture. In order to avoid the contamination, the samples
were kept in a flow of dry nitrogen during and after the polishing
until the end of the measurements. With this precaution, the
optical characteristics did not change noticeably during the
experiment. On the other hand, switching off the flow immediately
affected the optical signal as it is shown in the Appendix.

\section{Optical experiment and results}

Optical properties of MgB$_{2}$ at room temperature were obtained
using spectroscopic ellipsometry at 0.75 - 3.7 eV and the
reflectivity measurements in the infrared range from 0.1 to 0.85
eV. In all experiments samples were mounted on the top of a sharp
optically black cone and laser aligned.

The high-frequency spectra were collected using the Woollam VASE32
ellipsometer while the sample was kept in a flow of dry nitrogen.
The in-plane $\epsilon_{a}(\omega) = \epsilon_{1,a}(\omega) + 4\pi
i\sigma_{1,a}(\omega)/\omega$ and the c-axis $\epsilon_{c}(\omega)
= \epsilon_{1,c}(\omega) + 4\pi i\sigma_{1,c}(\omega)/\omega$
components of the complex dielectric tensor were both extracted
directly from the measurements on the ac-surface of the sample S1,
using two orthogonal crystal orientations and three angles of
incidence. For the sample S2, only the in-plane optical functions
$\epsilon_{1,a}(\omega)$ and $\sigma_{1,a}(\omega)$ were derived
from the measurement on the ab-plane, using the c-axis data from
the sample S1 to correct for the admixture of the out-of-plane
component. The details of the recovery of $\epsilon_{a}(\omega)$
and $\epsilon_{c}(\omega)$ from the ellipsometric output are given
in the Appendix.

\begin{figure}[bht]
\centerline{\includegraphics[width=8.5cm,clip=true]{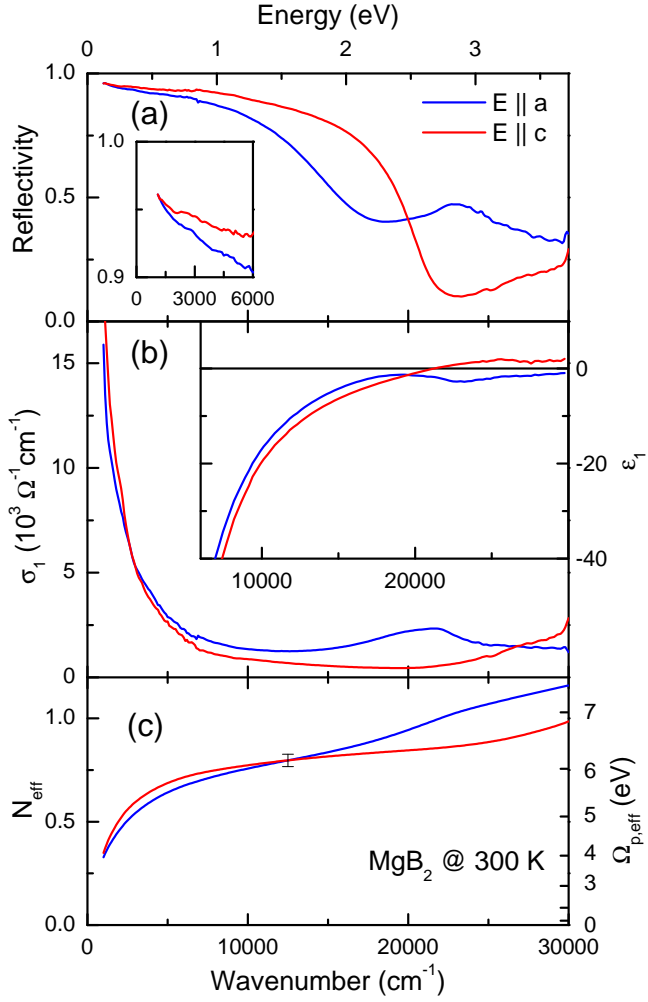}}
\caption{Optical anisotropic spectra of the MgB$_{2}$ at 300 K
derived from the ellipsometry and reflectivity measurements on the
sample S1 as described in the text: the normal-incidence
reflectivity $R(\omega)$(a), optical conductivity
$\sigma_{1}(\omega)$, the dielectric function
$\epsilon_{1}(\omega)$ (b), the effective number of carriers
$N_{\mbox{\scriptsize eff }}(\omega)$ and the effective plasma
frequency $\Omega_{\mbox{\scriptsize p,eff}}(\omega)$ (c). The
a-axis and the c-axis spectra are shown by the blue and red colors
respectively.}\label{FigDataAC}
\end{figure}

The reflectivity $R(\omega)$ was measured with a polarizer in the
range 0.1 - 0.85 eV at a near-normal angle of incidence using a
Fourier transform spectrometer Bruker 66v. The sample was inside a
vacuum chamber of a cryostat. A gold layer was sputtered {\it
in-situ} on the crystal surface to get a reference signal. We
reconstructed the full reflectivity spectrum using the
ellipsometrically determined dielectric functions in the optical
range:
\begin{equation}
R_{\nu}(\omega)=\left|\frac{1-\sqrt{\epsilon_{\nu}(\omega)}}{1+\sqrt{\epsilon_{\nu}(\omega)}}\right|^{2},
(\nu =a,c).
\end{equation}

The spectra from two regions were combined in order to obtain the
complex dielectric function in the whole range. While
$\epsilon(\omega)$ was measured directly above 0.75 eV, at low
frequencies we applied a variational Kramers-Kronig (KK)
constrained analysis of the data. In this method, described in
Ref.\onlinecite{KuzmenkoRSI05}, one finds a KK-consistent
dielectric function, which gives the best detailed match to the
reflectivity $R(\omega)$ at low frequencies and both
$\epsilon_{1}(\omega)$ and $\epsilon_{2}(\omega)$ at high
frequencies simultaneously. This procedure reduces to a large
extent the uncertainty due to extrapolations as compared to the
usual KK transform.

\begin{figure}[bht]
\centerline{\includegraphics[width=8.5cm,clip=true]{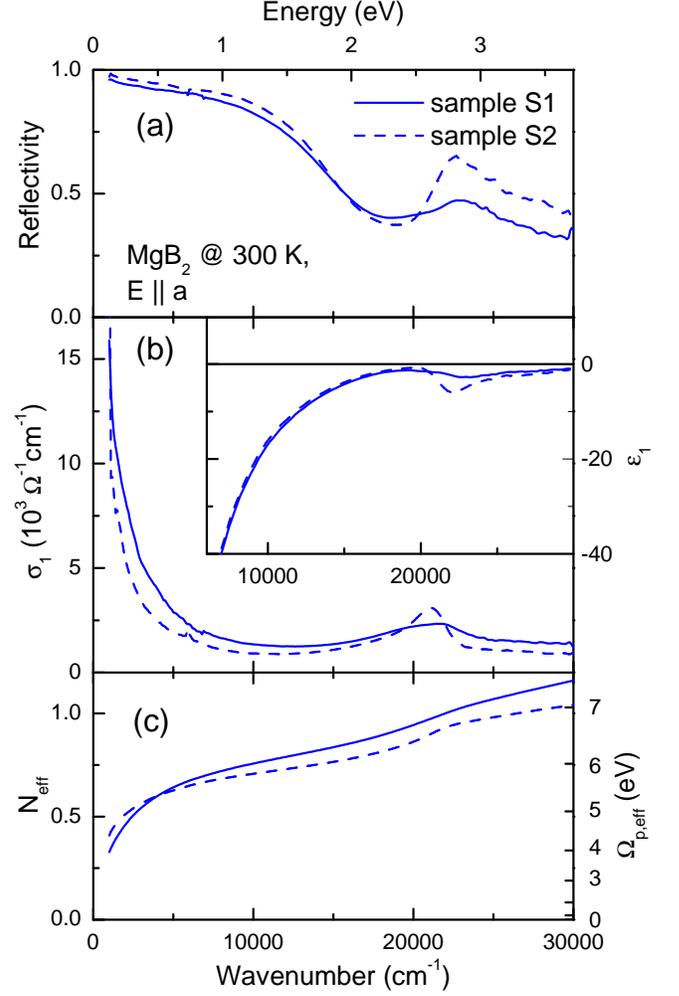}}
\caption{Comparison of the in-plane optical spectra measured on
different samples. The same types of spectra are presented as in
Fig.\ref{FigDataAC}. The spectra of samples S1 and S2 are shown by
the solid and dash lines, respectively. Sample S2 has presumably a
smaller concentration of impurities than sample
S1.}\label{FigDataS1S2}
\end{figure}

Figs.\ref{FigDataAC}a and \ref{FigDataAC}b show the anisotropic
reflectivity $R(\omega)$, optical conductivity
$\sigma_{1}(\omega)$ and the dielectric function
$\epsilon_{1}(\omega)$, measured on sample S1. Qualitatively, both
in-plane and c-axis spectra exhibit a similar metallic behavior,
characterized by a reflectivity plasma edge, a Drude peak in
$\sigma_{1}(\omega)$ and negative $\epsilon_{1}(\omega)$. However,
one can see a strong anisotropy. The in-plane reflectivity shows a
broad plasma edge at about 2 eV. The dielectric function
$\epsilon_{1,a}(\omega)$ almost reaches zero at the same
frequency, but then it varies non-monotonically while remaining
negative at least up to 3.7 eV. In contrast, the c-axis plasma
edge is significantly sharper and is higher in energy by about
0.5-0.6 eV. Correspondingly, $\epsilon_{1,c}(\omega)$ behaves
monotonically and crosses zero at $\omega_{p,c}^{\ast}\approx $
2.6 eV. These results agree with the previous findings of Fudamoto
and Lee \cite{FudamotoPRB03} who obtained a plasma edge at about
16000 cm$^{-1}$ (2 eV) on a mosaic of ab-oriented crystals.
Although no direct reflectivity measurements for $E\parallel c$
were reported so far, the same authors
suggested\cite{FudamotoPRB03} that the additional step-like
structure at 22000 cm$^{-1}$ (2.7 eV) in the reflectivity of
polycrystals samples comes from the c-axis plasma edge. The
present study fully confirms this assignment.

The in-plane optical conductivity $\sigma_{1,a}(\omega)$ shows an
intense interband peak at $\sim$ 2.6 eV, which is totaly absent in
the $\sigma_{1,c}(\omega)$ (Fig.\ref{FigDataAC}b). Its origin will
be discussed below. Since this peak is close to the screened
plasma frequency, it broadens the plasma edge and shifts it to
lower frequencies by providing an additional screening of the
charge carriers. It is also responsible for a non-monotonic
behavior of $\epsilon_{1,a}(\omega)$ at higher frequencies and
even for a structure in $R_{a}(\omega)$ above 3 eV, which
resembles a 'second' plasma edge.

Fig. \ref{FigDataAC}c depicts the partial sum rule (effective
number of carriers) function:
\begin{equation}\label{Neff}
N_{\mbox{\scriptsize eff }}(\omega) = \frac{2mV_{c}}{\pi
e^{2}}\int_{0}^{\omega}\sigma_{1}(\omega')d\omega',
\end{equation}
\noindent where $m$ is the free electron mass, $V_{c}$ = 28.9
\AA$^{3}$ is the unit cell volume, $e$ is the electron charge, and
the corresponding effective plasma frequency:
\begin{equation}
\Omega_{\mbox{\scriptsize p,eff}}(\omega)=
\left[8\int_{0}^{\omega}\sigma_{1}(\omega')d\omega'\right]^{1/2}.
\end{equation}

\noindent Since the integration of $\sigma_{1}(\omega)$ has to
start from zero frequency, while the optical data are taken down
to 100 meV, one has to be vigilant about the error bars involved.
Fortunately, the fact that the dielectric function obtained by the
method used\cite{KuzmenkoRSI05} satisfies the KK relations and
describes both $\epsilon_{1}(\omega)$ and $\epsilon_{2}(\omega)$
at high frequencies poses fairly tight bounds (shown in
Fig.\ref{FigDataAC}c) on the allowed values of
$N_{\mbox{\scriptsize eff }}$. One can see that at the photon
energy of 1.7 eV, which is high enough to comprise most of the
intraband spectral weight, but below the interband peak at 2.6 eV,
$N_{\mbox{\scriptsize eff }}$ is about 0.8 for both polarizations,
which corresponds to the plasma frequency of 6.3 eV. A more
rigorous estimate which takes into account the broadening of the
Drude peak, as described below, gives a very close value of
$\omega_{p}$. This is in contrast with the conclusion of
Ref.\onlinecite{FudamotoPRB03} about a strong anisotropy of the
plasma frequency, based on the different positions of the
reflectivity plasma edge for the two polarizations. The anisotropy
of the reflectivity plasma edge is caused by extra screening due
to the in-plane interband transition at 2.6 eV and not by the
strong anisotropy of the unscreened plasma frequency.

It is interesting to compare the in-plane optical properties of
the two samples (S1 and S2), which were prepared under different
conditions and have presumably slightly different impurity levels
(see Fig.\ref{FigDataS1S2}). One can see that the two most
prominent features of the optical conductivity - the Drude peak
and the 2.6 eV interband peak - are significantly sharper in the
sample S2. As a result, the reflectivity plasma edge is also
sharper and the 'double-plasmon' structure in $R_{a}(\omega)$ is
more pronounced than in sample S1. One should keep in mind that
the sample S2 was prepared from a slightly more pure magnesium,
and it shows a sharper superconducting transition
(Fig.\ref{FigTrans}). This tells that even a small impurity level
(about 0.2\% in this case) affects significantly optical and
transport properties.

The close values of $\epsilon_{1,a}(\omega)$ for the two samples
suggest that their in-plane plasma frequencies are similar,
although the $N_{\mbox{\scriptsize eff }}$ is slightly higher for
sample S1, which could be related to a stronger broadening of the
interband peaks.

\section{Discussion}

\subsection{Extended Drude analysis}\label{ExtendedDrude}

\begin{figure}[bht]
\centerline{\includegraphics[width=6cm,clip=true]{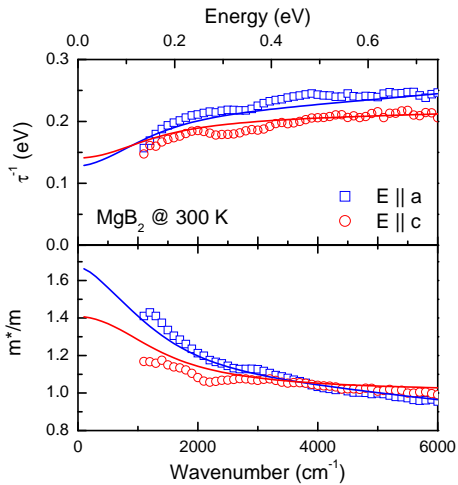}}
\caption{Extended Drude analysis of the optical conductivity of
MgB$_2$ (sample S1) at 300 K along the in-plane and the c-axis
directions. The symbols are the data, the solid curves show the
two-component fit as described in the Section
\ref{Fitting}.}\label{FigExtdrudeAC}
\end{figure}

The extended Drude model has been commonly used to analyze
interactions in electronic systems. In this formalism, the
frequency dependent scattering rate 1/$\tau(\omega)$ and effective
mass $m^{\ast}(\omega)/m$ are derived from the measured complex
dielectric function:
\begin{eqnarray}
\frac{1}{\tau(\omega)}=-\frac{\omega^{2}_{p}}{\omega}\mbox{ Im}\left(\frac{1}{\epsilon(\omega)-\tilde{\epsilon}_{\infty}}\right), \label{taui} \\
\frac{m^{\ast}(\omega)}{m}=-\frac{\omega^{2}_{p}}{\omega^{2}}\mbox{
Re}\left(\frac{1}{\epsilon(\omega)-\tilde{\epsilon}_{\infty}}\right).
\label{mstar}
\end{eqnarray}
The only parameters of this conversion are the total Drude plasma
frequency $\omega_{p}$ and the high-frequency dielectric constant
$\tilde{\epsilon}_{\infty}$, due to all contributions other than
the conductivity electrons. Both 1/$\tau(\omega)$ and
$m^{\ast}(\omega)/m$ have direct microscopic interpretation in the
context of the electron-phonon interaction. One should keep in
mind, however, that the model assumes that only one type of
carriers contributes to the Drude response. In the case of
MgB$_2$, which has two distinct systems of bands ($\sigma$ and
$\pi$), the scattering rate and the effective mass obtained by
Eqs.(\ref{taui}) and (\ref{mstar}) should be regarded as an
averaged value of contributions from each band.
Fig.\ref{FigExtdrudeAC} (symbols) shows 1/$\tau(\omega)$ and
$m^{\ast}(\omega)/m$ for the in-plane and the c-axis directions,
measured on the sample S1. We took the values $\omega_{p,a}$ =
6.28 eV, $\omega_{p,c}$ = 6.35 eV, $\tilde{\epsilon}_{\infty,a}$ =
11.9, $\tilde{\epsilon}_{\infty,c}$ = 4.77, which are suggested by
the data fitting in Section \ref{Fitting} (see Table
\ref{TableFit}). The value of $\tilde{\epsilon}_{\infty}$ here is
given by the sum of $\epsilon_{\infty}$ in the Table
\ref{TableFit} and the oscillator strengths $S_{i}$ of the Lorentz
oscillators. The solid curves in Fig.\ref{FigExtdrudeAC} were
calculated using the results of the fit described in Section
\ref{Fitting}.

One can see that the scattering rate and the mass renormalization
are larger for the in-plane direction. It is worth mentioning that
the $\sigma$-bands with a cylinder-like Fermi surface must have a
small electromagnetic response along the c-axis, while the
$\pi$-bands are expected to have comparable contributions in both
directions. The present result thus suggests that the
electron-phonon coupling is stronger in the $\sigma$-bands. This
is in agreement with the first-principle calculations of Kong {\it
et al.}\cite{KongPRB01}, who found that the total strength of the
electron-phonon interaction is dominated by the coupling of the
$\sigma$-holes to the bond-stretching optical phonons. A
quantitative analysis of the electron-phonon interaction must take
into account the multi-band electronic structure.

\subsection{Comparison with {\it ab initio} calculations}\label{Comparison}

A deeper insight can be attained by the comparison of the optical
data with the existing first-principle calculations of the band
structure and electron-phonon interaction. One can compute the
dielectric function, assuming that the intraband optical
conductivity is formed by the additive contributions of carriers
in the $\sigma$- and the $\pi$-bands. The two bands are
characterized by different electron-phonon (Eliashberg) transport
functions $\alpha^{2}_{tr}F(\omega)$ and impurity scattering rates
$\gamma_{imp}$. Intraband optical response can be calculated using
finite-temperature memory-function formalism for electron-boson
interaction \cite{DolgovJS05}. The total conductivity is a sum of
the intraband, interband ($\epsilon^{IB}$) and core-electron
($\epsilon_{\infty}$) responses:
\begin{eqnarray}\label{intraband}
\epsilon_{\nu}(\omega) =
\sum_{\beta=\sigma,\pi}\frac{\omega_{p,\nu\beta}^{2}}{-\omega[\omega+iM_{\beta}(\omega,T)]}+
\epsilon_{\nu}^{IB}(\omega)+\epsilon_{\nu\infty}\\ \nonumber (\nu
= a,c).
\end{eqnarray}

\noindent The contribution of each of the two bands is determined
by its respective anisotropic plasma frequency ($\omega_{p,a}$,
$\omega_{p,c}$) and the memory function\cite{DolgovJS05}:
\begin{equation}\label{memory}
M_{\beta}(\omega,T) = \gamma_{\beta
imp}-2i\int_{0}^{\infty}d\Omega\
\alpha^{2}_{tr}F_{\beta}(\Omega)K\left(\frac{\omega}{2\pi
T},\frac{\Omega}{2\pi T}\right),
\end{equation}

\noindent where
\begin{eqnarray}\label{}
K(x,y) = \frac{i}{y} &+& \frac{y-x}{x}[\psi(1-ix+iy)-\psi(1+iy)]+\nonumber\\
& & \frac{y+x}{x}[\psi(1-ix-iy)-\psi(1-iy)],\nonumber
\end{eqnarray}

\noindent and $\psi(x)$ is a digamma function.

All the ingredients to compute $\epsilon(\omega)$, except the
impurity scattering rates $\gamma_{\sigma imp}$ and $\gamma_{\pi
imp}$, which depend on the level and the nature of the impurities
(substitutions, vacancies, dislocations etc.), are provided by the
{\it ab initio} LDA calculations.

The spectra of the interband optical conductivity
$\sigma_{1}^{IB}(\omega)$ obtained by different groups using the
LMTO\cite{AntropovCM01,RavindranPRB01}, full-potential
LAPW\cite{KortusPRL01} and time-dependent DFT\cite{KuPRL02}
methods, although showing some differences, are close to each
other. In particular, we took the interband conductivity spectra,
presented in Ref.\onlinecite{KuzmenkoSSC02} (Fig.5) on the base of
the full-potential LAPW method\cite{KortusPRL01}, and computed the
corresponding dielectric function $\epsilon_{1}^{IB}(\omega)$ by
the KK transformation. The plasma frequencies reported by
different groups \cite{KortusPRL01,AntropovCM01,RavindranPRB01}
are also close to each other. We used the values
$\omega_{p,a\sigma}$ = 4.14 eV, $\omega_{p,a\pi}$ = 5.89 eV,
$\omega_{p,c\sigma}$ = 0.68 eV and $\omega_{p,c\pi}$ = 6.85
eV\cite{KortusPRL01,MazinPRL02}.

The {\it ab initio} calculated electron-phonon interaction
functions were presented in Ref.\onlinecite{KongPRB01}. We used
effective interaction functions for the two bands
$\alpha^{2}_{tr}F_{\sigma}(\omega) \equiv
\alpha^{2}_{tr}F_{\sigma\sigma}(\omega) +
\alpha^{2}_{tr}F_{\sigma\pi}(\omega)$ and
$\alpha^{2}_{tr}F_{\pi}(\omega) \equiv
\alpha^{2}_{tr}F_{\pi\pi}(\omega) +
\alpha^{2}_{tr}F_{\pi\sigma}(\omega)$ from the
Refs.\onlinecite{KongPRB01,GolubovJPhysCM02,DolgovCommunication}.
The electron-phonon coupling constant $\lambda_{tr} =
2\int_{0}^{\infty}d\Omega\alpha^2_{tr}F(\Omega)/\Omega$ is
calculated to be about 1.1 for the $\sigma$-bands and 0.55 for the
$\pi$-bands.

\begin{figure}[bht]
\centerline{\includegraphics[width=8.5cm,clip=true]{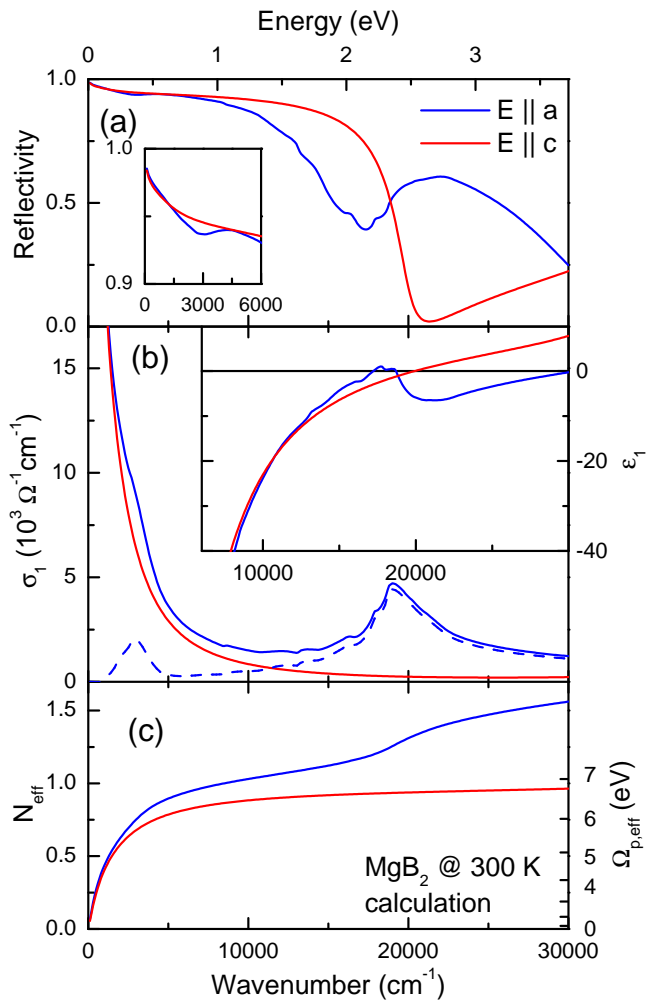}}
\caption{Anisotropic optical spectra of MgB$_{2}$ at 300 K,
calculated using the results of the published first-principle
studies\cite{KortusPRL01,KongPRB01} as described in the text. The
same types of spectra are shown as in Fig.\ref{FigDataAC}.
Additionally, the interband contribution along the a-axis is
presented in (b) as a dashed blue line. The interband conductivity
along the c-axis is negligible in the shown energy
range.}\label{FigTheoryAC}
\end{figure}

The calculated optical spectra are presented in
Fig.\ref{FigTheoryAC}. Here the scattering rates $\gamma_{\sigma
imp}$ = 12.4 meV and $\gamma_{\pi imp}$ = 85.6 meV were taken;
this choice is substantiated by the data fitting in the next
Section. Comparing Figs.\ref{FigDataAC} and \ref{FigTheoryAC}, one
can notice a remarkable overall agreement and even a reasonable
quantitative match.

We begin with the interband transitions. Both theoretical and
experimental interband optical conductivities show a very strong
anisotropy. Notably, the theory predicts no sizeable optical
intensity along the c-axis of the interband transitions below 4
eV, whereas there are two low-lying peaks at 0.35 eV and 2.4 eV
for polarization parallel to the boron planes (see the blue dashed
curve in Fig.\ref{FigTheoryAC}b). The 2.4 eV peak is due to a
transition from the $\sigma$-band to the $\pi$-band close to the
M-point, where a van Hove singularity strongly enhances the
density of states \cite{KortusPRL01}. The peak at 0.35 eV
corresponds to the transition between two close $\sigma$-bands.
The $\sigma$ carriers are localized within the boron planes, which
explains why these two transitions are optically very weak along
the c-axis.

Obviously, a strong peak in the experimental spectrum along the
a-axis at 2.6 eV matches the theoretical peak at 2.4 eV. The
discrepancy in the peak position suggests that the separation
between $\sigma$-band and $\pi$-band is about 0.2 eV bigger than
predicted by the theory. It is remarkable\cite{RosnerPRB02}, that
the same shift brings the results of the dHvA experiments
\cite{YellandPRL02} in MgB$_2$ close to theoretical
predictions\cite{RosnerPRB02,MazinPRB02}.

According to the calculation, the $\sigma\rightarrow\sigma$
transition at 0.35 eV should manifest itself as a noticeable dip
in the reflectivity spectrum (see inset in
Fig.\ref{FigTheoryAC}a). Such a dip is not observed in the
experimental spectra (Fig.\ref{FigDataAC}a). This peak may thus be
shifted to even lower frequencies or heavily overdamped. The first
possibility is consistent with a shift of the $\sigma$-bands with
respect to the Fermi energy compared to the calculation, as
proposed in Ref.\onlinecite{RosnerPRB02}. The broadening can be
caused by the interaction with the conduction electrons, since the
energy of this transition is within the width of the Drude peak.

A very intense sharp peak in the out-of-plane conductivity
$\sigma_{1,c}(\omega)$ is expected around 5 eV
\cite{AntropovCM01,KuzmenkoSSC02,KuPRL02}. This excitation was
closely studied in Ref.\onlinecite{KuPRL02} by the time-dependent
DFT method. In the band picture, it comes from a transition
between almost parallel bands. From the real-space point of view,
this mode involves charge fluctuations between B and Mg sheets,
dynamically screened by the intraband transitions\cite{KuPRL02}.
As a consequence, a sharp plasma mode at 2.5 eV should emerge.
Although 5 eV is beyond our experimental range, there is a
sizeable increase of $\sigma_{1,c}(\omega)$ above 3 eV, which can
be a low-energy tail of this mode. The c-axis plasmon is at 2.6
eV, which is only slightly higher than the calculated value
($\sim$ 2.5 eV). This is in agreement with the inelastic X-ray
scattering experiment\cite{GalambosiPRB05}.

The shape and the width of the Drude peak agree well with the
experiment, which suggests that phonons and impurities are the
main factors of electron scattering. The integrated spectral
weight $N_{\mbox{\scriptsize eff }}(\omega)$ grows faster
according to the calculations, especially for the in plane
response. This indicates that the theory overestimates the value
of the plasma frequency.

\subsection{Two-band fitting of the spectra}\label{Fitting}

\begin{figure}[bht]
\centerline{\includegraphics[width=6cm,clip=true]{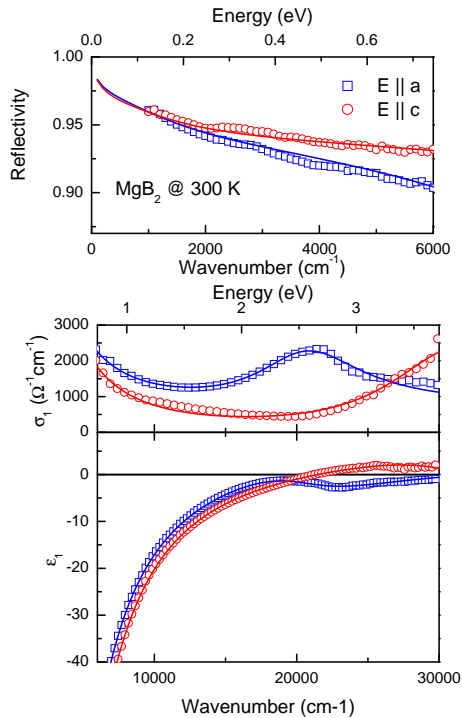}}
\caption{Experimental data (symbols) and the multi-band fit (solid
lines) of the reflectivity (a) and the dielectric function (b)
along the a-axis (blue) and the c-axis (red).} \label{FigFit}
\end{figure}

The overall good agreement between the experimental and calculated
data allows us to use the chosen model to fit the spectra,
treating the plasma frequencies and scattering rates as adjustable
parameters. We could achieve a satisfactory least-square fit of
reflectivity $R(\omega)$ in the mid-infrared and
$\epsilon_{1}(\omega)$ and $\sigma_{1}(\omega)$ at higher
frequencies using formulas (\ref{intraband}) and (\ref{memory}) as
it is shown in Fig.\ref{FigFit}. The corresponding parameter
values are given in Table \ref{TableFit}.

It turns out that leaving all four plasma frequencies adjustable
makes the fitting procedure under-determined. Therefore, assuming
the 2D nature of the $\sigma$-band, we fixed the plasma
frequencies of the $\sigma$-band to the values given by {\it ab
initio} calculations, and left only the total in-plane and c-axis
plasma frequencies $\omega_{p,tot}^2 = \omega_{p,\sigma}^2 +
\omega_{p,\pi}^2$ adjustable. Another assumption was that the {\it
ab initio}
calculations\cite{KongPRB01,GolubovJPhysCM02,DolgovCommunication}
correctly describe the electron-phonon functions
$\alpha^{2}_{tr}F_{\sigma,\pi}(\omega)$.

We modelled the interband conductivities by Lorentz oscillators
($\nu = a,c$):
\begin{equation}
\epsilon_{\nu}^{IB}(\omega) = \sum_{i}\frac{S_{\nu i}\omega_{\nu
i}^{2}}{\omega_{\nu i}^{2} - \omega^{2}-\gamma_{\nu i}\omega}.
\end{equation}

\noindent with adjustable frequency $\omega_{\nu i}$, oscillator
strength $S_{\nu i}$ and width $\gamma_{\nu i}$. Keeping in mind
the two interband peaks below 3 eV predicted by the theory, we put
two Lorentzians to model $\epsilon_{a}^{IB}(\omega)$. Only one
oscillator term above 3 eV was taken for
$\epsilon_{c}^{IB}(\omega)$.

One can see that the bare plasma frequencies $\omega_{p,\sigma}$
and $\omega_{p,\pi}$ are almost the same and equal to 6.3 eV,
which confirms our previous estimate based on the partial sum rule
(Fig.\ref{FigDataAC}c). This value is much higher compared to
previous
reports\cite{TuPRL01,KaindlPRL02,KuzmenkoSSC02,MazinPRL02,MunJS02,FudamotoPRB03}
of 1.5-2.5 eV. Thus, the discrepancy with the theoretical value of
7 eV is likely to be much less than it was thought before.
However, the current mismatch is not negligible since it results
in about 20-25\% deviation of the Drude spectral weight. It is
worth to mention that the extremal orbit areas in the de Haas-van
Alphen experiment\cite{YellandPRL02} on both $\sigma$ and $\pi$
Fermi surfaces are also somewhat smaller than predicted by theory
\cite{MazinPRB02,RosnerPRB02}. It was pointed out
\cite{RosnerPRB02} that the discrepancy can be removed by a shift
of the $\sigma$-bands downward by about 115 meV and the
$\pi$-bands upward by 125 meV. This is in perfect agreement with
our observation of the mismatch of 0.2 eV in the position of the
2.6 eV peak, as it is mentioned before. This also qualitatively
explains the smaller value of the plasma frequency.

\begin{figure}[bht]
\centerline{\includegraphics[width=8.5cm,clip=true]{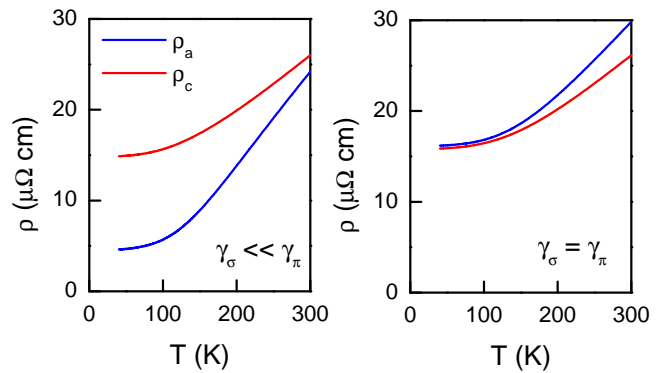}}
\caption{Temperature-dependent resistivity curves along the a-axis
(blue) and c-axis (red), calculated using the results of the fit
of the optical data for $\gamma_{\pi imp}$ = 85.6 meV and
different values of the scattering rate in the $\sigma$-band:
$\gamma_{\sigma imp}$ = 12.4 meV $\ll\gamma_{\pi imp}$(left) and
$\gamma_{\sigma imp} = \gamma_{\pi imp}$ (right).
}\label{FigResistivity}
\end{figure}

The impurity scattering rate in the $\pi$-band $\gamma_{\pi imp}$
is about 85 meV. Since the $\pi$-electrons have rather strong
optical spectral weight and a modest electron-phonon interaction,
the absolute values of both $R_{a}(\omega)$ and $R_{c}(\omega)$
are quite sensitive to this parameter. In contrast, the value of
$\gamma_{\sigma imp}$ does not significantly affect the spectra
and cannot be accurately determined from the fit, because a large
electron-phonon interaction in the $\sigma$-band dominates the
total scattering above 100 meV at 300 K. However, $\gamma_{\sigma
imp}$ influences drastically the shape of the
temperature-dependent in-plane DC resistivity
$\rho_{a}(T)=1/\sigma_{1,a}(\omega\rightarrow 0,T)$ curves, most
notably the value of $\rho(300 K)/\rho(40 K)$ (RRR), as can be
seen from Fig.\ref{FigResistivity}. A measurement
\cite{SologubenkoPRB02} on a crystal grown in the same group under
similar condition, showed RRR $\approx$ 5 along the a-axis.
Therefore we have chosen of $\gamma_{\sigma imp}$ = 12.4 meV,
which gives the same resistivity ratio (see
Fig.\ref{FigResistivity}). The assumption $\gamma_{\sigma imp} =
\gamma_{\pi imp}$ gives RRR less than two, which strongly suggests
that $\gamma_{\sigma imp}$ is actually several times smaller than
$\gamma_{\pi imp}$. This agrees with the Raman study
\cite{QuiltyPRL03} on single crystals, where a relation
$\gamma_{\pi imp}/\gamma_{\sigma imp} = 6 - 9$ was found, although
the absolute values of the scattering rates are smaller than in
our case.

\begin{table}
\caption{Model parameters which give the best match to the
measured optical spectra of MgB$_{2}$ at 300 K and the ratio
$\rho_{a}(300 K)/\rho_{a}(40 K)$ from
Ref.\onlinecite{SologubenkoPRB02}, as described in the text.
Parameters, marked by ${\ast}$, were fixed to the values given by
the band calculations.}
\begin{tabular}{cccc}
  \hline
  \hline
  Parameter &ab-plane & c-axis\\
  \hline
  $\omega_{p,\sigma}$ & 4.14$\ast$ eV & 0.68$\ast$ eV \\
  $\omega_{p,\pi}$    & 4.72 eV & 6.31 eV\\
  $\omega_{p,tot}$    & 6.28 eV & 6.35 eV\\
  $\gamma_{\sigma imp}$   &\multicolumn{2}{c}{12.4 meV}\\
  $\gamma_{\pi imp}$      &\multicolumn{2}{c}{85.6 meV} \\
  \hline
  $\omega_{1}$      & 2.60 eV & 3.92 eV\\
  $S_{1}$             & 1.79 & 1.7\\
  $\gamma_{1}$        & 1.11 eV & 1.51 eV\\
  $\omega_{2}$      & 8.57 eV & - \\
  $S_{2}$             & 6.81  & - \\
  $\gamma_{2}$        & 100.2 eV & - \\
  \hline
  $\epsilon_{\infty}$   & 3.31 & 3.07 \\
  \hline
  \hline
\end{tabular}
\label{TableFit}

\end{table}

A disparity between the impurity scattering rates in the two bands
and a small interband $\sigma\leftrightarrow\pi$ scattering were
proposed in Ref.\onlinecite{MazinPRL02} to explain the
surprisingly small dependence of $T_{c}$ on the impurity level,
not expected for a two-band superconductor. Microscopically, this
can be explained by the fact that the electronic wavefunctions in
$\sigma$-bands are confinedto the boron planes and not efficiently
scattered by the magnesium vacancies and substitutions. On the
other hand, the same defects strongly scatter the electron states
in the 3D $\pi$-bands. Our results support this conjecture.

The first Lorentz term for the a-axis at 2.6 eV clearly
corresponds to the discussed already interband transition.
However, the second oscillator is extremely broad and it cannot be
matched with a predicted interband peak at 0.35 eV. This probably
means this relatively weak peak is shifted and/or broadened so
much that it cannot be identified in the optical spectrum. The
second Lorentz term may thus describe the background formed by the
tails of several broadened interband peaks.

\subsection{The color of the magnesium diboride}

\begin{figure}[btp]
\centerline{
\includegraphics[width=8cm,clip=true]{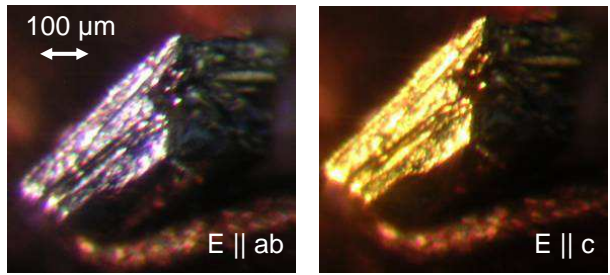}}
\caption{Two images of the same sample (ac-plane) made in a
polarized light: $E\parallel ab$ (left) and $E\parallel c$
(right). The sample looks golden for $E\parallel c$, due to a
sharp reflectivity plasma edge at around 2.5 eV, while the plasma
edge is smeared for $E\parallel ab$ by a strong
$\sigma\rightarrow\pi$ interband transition at 2.6 eV.}
\label{FigColor}
\end{figure}

Finally we address a practical question: what is the color of
MgB$_{2}$? The characterizations given in the literature varied
significantly, ranging from the 'golden' and 'tan' to
'silver-metallic', 'black' and even 'blue'. Lee \cite{LeePC03} has
noticed that, depending on the polarization, "...the fresh surface
of as-grown crystals in the ab-plane changes the color from silver
to dark gray. In contrast, for the ac-plane of the crystals a
beautiful change of color from golden-yellow to blue can be seen."
Fig.\ref{FigColor} shows the two images of the unpolished ac-plane
of the sample S1 for different polarizations of the incident white
light. One can see that the sample spectacularly changes color
from a blueish-silver for $E \parallel ab$ to the yellow for $E
\parallel c$. The explanation comes naturally from the
reflectivity curves shown in Fig.\ref{FigDataAC}a. For the c-axis
polarization, it is a sharp plasma edge at about 2.5 eV, the same
as of gold, which makes the sample yellow. For the ab-plane
polarization, the plasma edge is smeared due to the strong
interband transition at 2.6 eV. As a result, the reflectivity
spectrum in the visible range (1.8 - 3.1 eV) is relatively flat
with a maximum (stronger in sample S2) at $\sim$ 2.8 eV which
makes the color blue-silver.

\begin{figure}[bht]
\centerline{\includegraphics[width=8cm,clip=true]{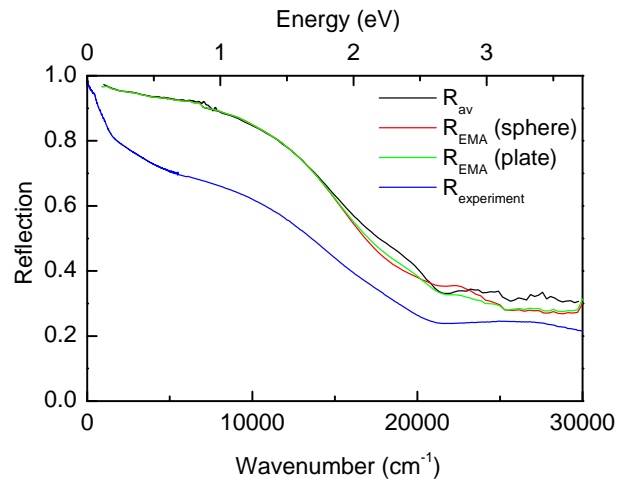}}
\caption{Reflectivity spectra of polycrystalline MgB$_{2}$ at 300
K. The spectra, calculated using the effective-medium
approximation (spherical or plate-like crystallites) and the
direct reflectivity averaging, are compared with experimental
spectrum from Ref.\onlinecite{KuzmenkoSSC02}.}
\label{FigPolycryst}
\end{figure}

Knowing the anisotropic spectra, we can predict the reflectivity
shape of MgB$_{2}$ in the polycrystalline form. If the light
wavelength $\lambda$ is much larger than a typical grain size $l$,
one can apply the effective-medium approximation (EMA) as
prescribed in Ref.\onlinecite{SulewskiPRB87}. In the
short-wavelength limit $\lambda \ll l$, it is more adequate to
average directly the reflectivities along the two axes, as it was
done in Ref.\onlinecite{FudamotoPRB03}. In Fig.\ref{FigPolycryst}
both calculated curves $R_{EMA}(\omega)$ (for sphere-like and
plate-like crystallites) and
$R_{av}(\omega)=(2/3)R_{ab}(\omega)+(1/3)R_{c}(\omega)$ are shown,
together with the spectrum on a polycrystal from our previous
study \cite{KuzmenkoSSC02}. Notably, $R_{EMA}(\omega)$ and
$R_{av}(\omega)$ are close to each other and their overall shape
in the visible range matches well the measured spectrum. The
latter shows only a weak step-like feature near 2.5 eV, coming
from the $c$-axis contribution\cite{FudamotoPRB03}, which explains
why the polycrystalline samples have typically black, or slightly
tan, color. The absolute value of the experimental reflectivity
from Ref.\onlinecite{KuzmenkoSSC02} is by 10-20\% lower than the
calculated one, which is likely caused by an overdamping of the
$\pi$-bands due to a much higher impurity level (presumably MgO)
of the sample used.

\section{Conclusions}

We investigated anisotropic optical properties of MgB$_2$ single
crystals of different purity at room temperature in the energy
range 0.1 - 3.7 eV and compared them with the existing first
principle calculations of the electronic structure and
electron-phonon coupling. The analysis of the anisotropy allowed
us to distinguish properties of the $\sigma$- and $\pi$-bands.

The total bare plasma frequencies along the in-plane and c-axis
directions are almost the same and equal to 6.3 eV, which is much
higher than the previously reported values. However, it is still
smaller than the theoretical value of about 7 eV.

The shape of the Drude peak is well described by the electron
scattering on phonons and impurities. The data suggest that the
impurity scattering in the $\pi$-bands is several times larger
than in the $\sigma$-bands in agreement with a proposal of
Ref.\onlinecite{MazinPRL02}, aimed to explain a surprisingly small
dependence of the critical temperature on the sample purity. The
electron-phonon interaction is stronger in the $\sigma$-bands, in
agreement with the theory\cite{KongPRB01} and de Haas-van Alphen
experiments\cite{YellandPRL02}.

The in-plane optical conductivity clearly shows an intense peak at
2.6 eV, which corresponds to a transition from the $\sigma$- to
the $\pi$-band. It is higher by about 0.2 eV than the theoretical
value. An interband peak due to the transitions between two close
$\sigma$-bands, expected to be at 0.35 eV, was not seen in the
experiments which can be explained by an overdamping or a shift
towards lower frequencies relative to the theoretical
calculations.

The width of the Drude peak and the interband
$\sigma\rightarrow\pi$ peak at 2.6 eV are very sensitive to the
sample purity. This can partially explain the spread of the
results of previous optical studies.

The two quantitative mismatches between the experiment and the
theory, namely (i) a smaller plasma frequency, (ii) a higher by
0.2 eV energy of the 2.6 eV $\sigma\rightarrow\pi$ transition,
tell us that the $\sigma$-bands are probably shifted down and
$\pi$-bands are shifted up compared to the calculations, so that
their relative shift is about 0.2 eV. Notably, the same conclusion
was derived in Ref.\onlinecite{RosnerPRB02} from the analysis of
the dHvA measurements\cite{YellandPRL02}.

The positions of the reflectivity plasma edges for two
polarizations are very different ($\sim$ 2 eV for a-axis and
$\sim$ 2.5 eV for the c-axis). This is caused by an additional
screening of the intraband carriers by the interband transition at
2.6 eV, but not due to different ab-plane and c-axis bare plasma
frequencies. As a result, the color of the sample depends on the
polarization of the light: it is blueish-silver for E $\parallel$
a and yellow for E $\parallel$ c.

This work was supported by the Swiss National Science Foundation
through the National Center of Competence in Research "Materials
with Novel Electronic Properties - MaNEP". We are grateful to O.V.
Dolgov, J. Kortus and I.I. Mazin for fruitful discussions and
kindly sharing their computational data. We thank R. Tediosi for
making a picture of the sample.
\newpage

\section*{Appendix. Determination of $\epsilon_{a}(\omega)$ and
$\epsilon_{c}(\omega)$ from ellipsometry}

\begin{figure}[bht]
\centerline{\includegraphics[width=6cm,clip=true]{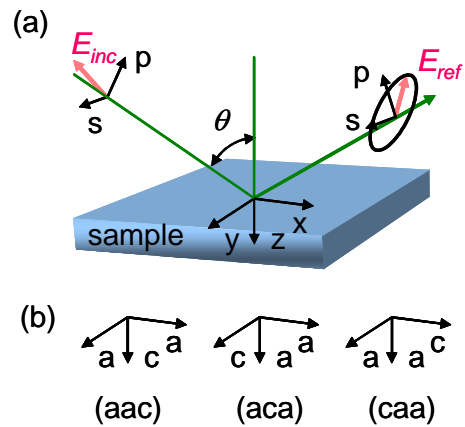}}
\caption{(a) The general configuration of ellipsometric
experiment; (b) three specific sample orientations used in this
paper.} \label{FigGeometries}
\end{figure}

The general geometry of the ellipsometric experiment in the
configuration 'fixed analyzer (at 45$^{\circ}$) - sample -
rotating analyzer' is shown on Fig.\ref{FigGeometries}a. The axes
$x$, $y$ and $z$ are assumed to be along the principal axes of the
dielectric tensor of the sample. Ellipsometry provides two
parameters $\psi$ and $\Delta$, related to the ratio of reflection
coefficients for the $p$- and $s$-polarized light:
\begin{equation}
\rho = r_{p}/r_{s} = \tan{\psi} \exp(i\Delta).
\end{equation}
\noindent $r_{p}$ and $r_{s}$ are given by the Fresnel formulas:
\begin{eqnarray}
r_{p} &=&
\frac{\sqrt{1-\varepsilon_{z}^{-1}\sin^{2}\theta}-\sqrt{\varepsilon_{x}}\cos
\theta}
{\sqrt{1-\varepsilon_{z}^{-1}\sin^{2}\theta}+ \sqrt{\varepsilon_{x}}\cos \theta}, \label{rp} \\
r_{s} &=& \frac{\cos \theta-\sqrt{\varepsilon_{y}-\sin^{2}\theta}}
{\cos\theta + \sqrt{\varepsilon_{y}-\sin^{2}\theta}},\label{rs}
\end{eqnarray}

\noindent where $\epsilon_{\nu} = \epsilon_{1\nu} +
i\epsilon_{2\nu}$ ($\nu$ = $x$, $y$, $z$) are the components of
the complex dielectric tensor. In the case of optically uniaxial
MgB$_{2}$ ($\epsilon_{a}=\epsilon_{b}\neq\epsilon_{c}$) three
different orientations ($xyz$) are possible: ($aac$), ($aca$) and
($caa$) (see Fig.\ref{FigGeometries}b).

\begin{figure}[bht]
\centerline{\includegraphics[width=7cm,clip=true]{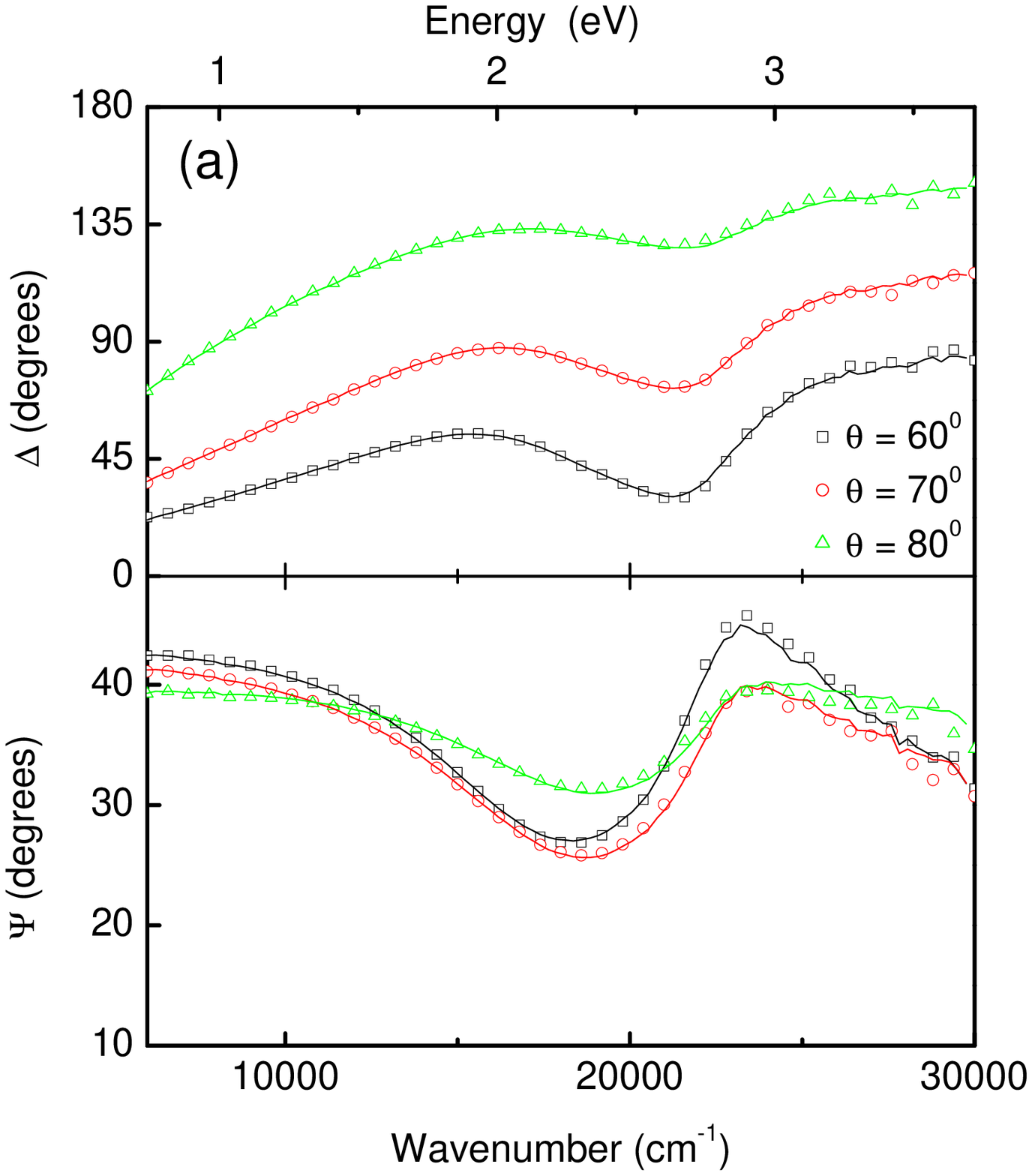}}
\centerline{\includegraphics[width=7cm,clip=true]{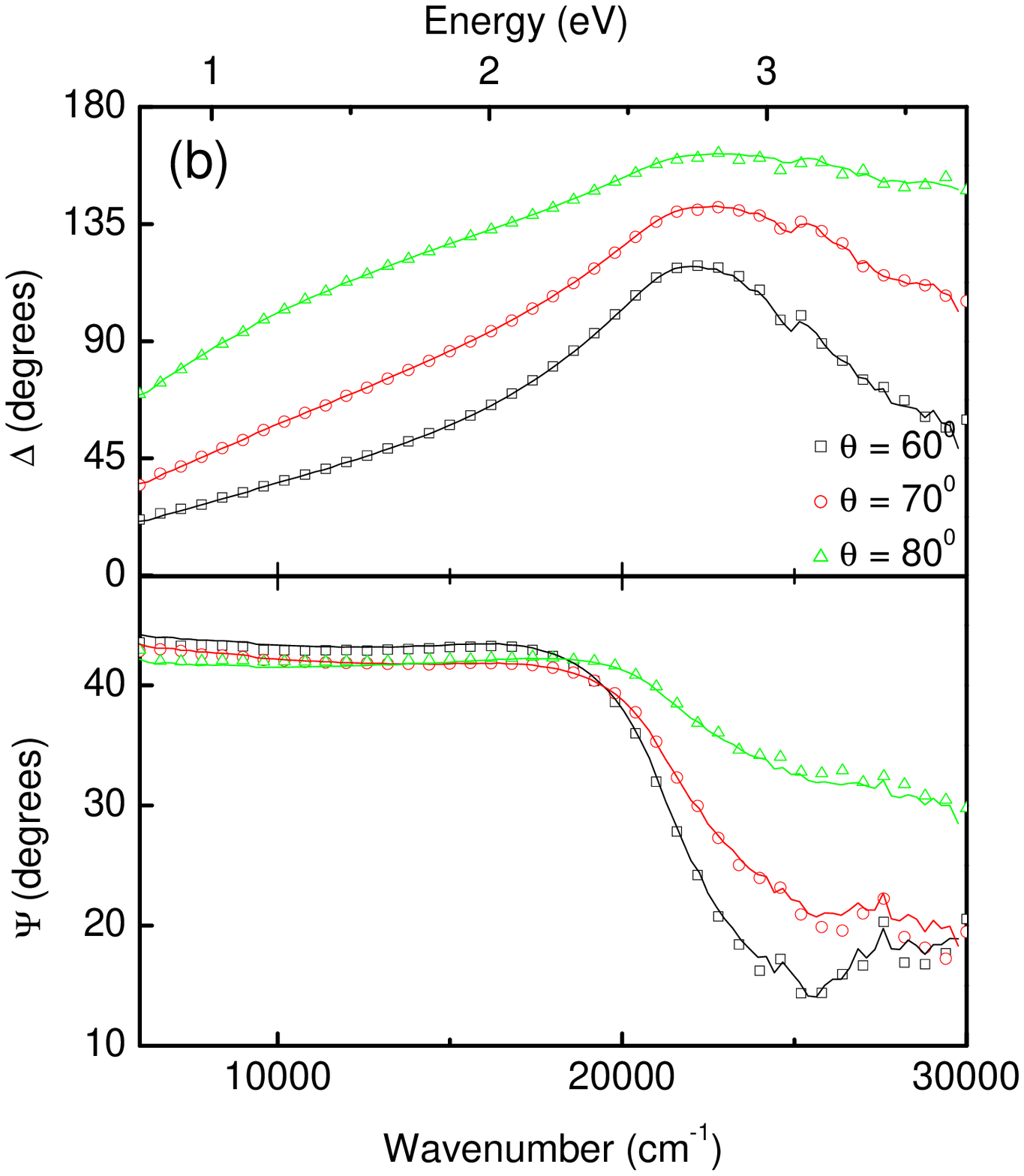}}
\caption{Ellipsometric spectra $\psi$ and $\Delta$ for three
angles of incidence (60$^{\circ}$, 70$^{\circ}$, and 80$^{\circ}$)
in the orientations ($aca$)(a) ($caa$)(b) and (see
Fig.\ref{FigGeometries}b), measured on sample S1. Symbols are the
measurement results, the solid line correspond to the fit as
described in the text.} \label{FigPsiDelS1}
\end{figure}

On the sample S1 we did ellipsometry in the orientations ($caa$)
and ($aca$). This yields four independent quantities $\psi_{aca}$,
$\Delta_{aca}$, $\psi_{caa}$ and $\Delta_{caa}$ at every photon
energy and every chosen angle of incidence. Since each of these
functions depend on the four values $\epsilon_{1,a}$,
$\epsilon_{2,a}$, $\epsilon_{1,c}$ and $\epsilon_{2,c}$, the
latter ones can be obtained by the numerical inversion of the four
Fresnel equations. This procedure, when applied to three different
angles of incidence (60$^{\circ}$, 70$^{\circ}$ and $80^{\circ}$),
gave close results. In order to improve the accuracy of the
output, we determined by the least square fitting the values of
$\epsilon_{1,a}$, $\epsilon_{2,a}$, $\epsilon_{1,c}$ and
$\epsilon_{2,c}$ that render the best match to the measured
ellipsometric parameters at all mentioned angles of incidence
simultaneously. The experimental and fitting curves for the sample
S1 for the orientations ($aca$) and ($caa$) are shown in
Figs.\ref{FigPsiDelS1}a and \ref{FigPsiDelS1}b respectively.

\begin{figure}[bht]
\centerline{\includegraphics[width=7cm,clip=true]{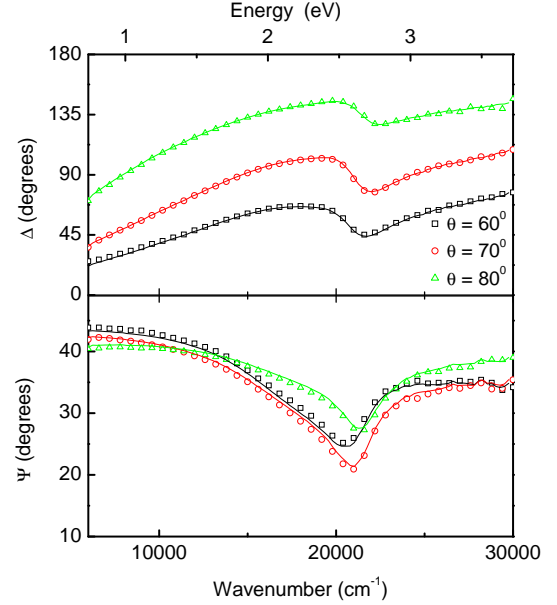}}
\caption{Ellipsometric spectra $\psi$ and $\Delta$ for three
angles of incidence (60$^{\circ}$, 70$^{\circ}$, and 80$^{\circ}$)
in the orientation ($aac$) (see Fig.\ref{FigGeometries}b),
measured on sample S2. Symbols are the measurement results, the
solid line correspond to the fit as described in the text.}
\label{FigPsiDelS2}
\end{figure}

The second experiment was done on the sample S2 in the orientation
($aac$). The corresponding spectra of $\psi_{aac}$ and
$\Delta_{aac}$ for the same three angles of incidence are shown in
Fig.\ref{FigPsiDelS2} (as solid symbols). These data, taken alone,
are not sufficient to extract both $\epsilon_{a}$ and
$\epsilon_{c}$ independently. Since $\psi_{aac}$ and
$\Delta_{aac}$ are not very sensitive to the value of
$\epsilon_{c}$ except close to the screened plasma frequency
(according to Aspnes approximation) \cite{AspnesJOSA80}, we
extracted $\epsilon_{a}(\omega)$, assuming that
$\epsilon_{c}(\omega)$ is the same as the sample S1. One should
realize that this introduces some uncertainty, because the samples
S1 and S2 might have somewhat different c-axis dielectric
functions. This uncertainty is negligible below 2 eV, but somewhat
enhanced close to the c-axis plasmon at 2.6 eV. The best fit of
$\psi$ and $\Delta$ is given by the solid lines in
Fig.\ref{FigPsiDelS2}.

\begin{figure}[bht]
\centerline{\includegraphics[width=5cm,clip=true]{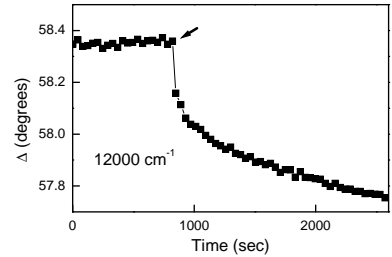}}
\caption{The effect of the exposure to the air on the optical
properties of MgB$_2$. The time dependence of ellipsometric
parameter $\Delta$ at 1.5 eV is shown (sample S1, geometry (caa),
angle of incidence 70$^\circ$). Initially the sample was kept in a
flow of dry nitrogen, which was switched off at the moment
designated by the arrow.} \label{FigDeter}.
\end{figure}

Fig.\ref{FigDeter} shows how quickly the ellipsometric parameters
change when a surface is exposed to the air, due to the formation
of a contamination layer.

\end{document}